
\magnification=1200 \def\wc{\hangindent=4em \hangafter=1 \noindent}
\baselineskip 14pt \parskip 3pt \null 
\headline={\ifnum\pageno=1\hfil\else\hfil\tenrm--\ \folio\ --\hfil\fi}
\footline={\hfil}

\centerline{\bf The Masses of Nearby Dwarfs can be Determined}
\centerline{\bf with Gravitational Microlensing}
\vskip 0.5cm
\centerline{by}
\vskip 0.5cm
\centerline{B. Paczy\'nski}
\vskip 0.5cm
\centerline{Princeton University Observatory, Peyton Hall, Princeton,
NJ 08544-1001}
\centerline{Visiting Scientist, National Astronomical Observatory, Mitaka,
Tokyo, 181, Japan}
\centerline{e-mail: bp@astro.princeton.edu}

\vskip 0.5cm
\centerline{\it Received: .........................}
\vskip 0.5cm
\centerline {ABSTRACT}
\vskip 0.5cm
Microlensing of distant stars in the Milky Way by the nearby high proper
motion stars offers a direct way to precisely measure the masses of single
lower main sequence stars and brown dwarfs.


\vskip 0.5cm

Gravitational microlensing can be used to determine the masses of lensing
objects, as the time scale of the event is proportional to the
square root of the lensing mass.  Unfortunately, the time scale also depends
on the distance to the lens and the source and their relative transverse
motion, making the mass determination rather uncertain (cf. Paczy\'nski 1991
and references therein).  In a special case of compact masses located in a
globular cluster lensing distant stars the mass determination may be more
robust, but the expected rate of events is very low (Paczy\'nski 1994).
In this paper another possibility of accurate determination of microlensing
masses is presented.

Every high proper motion star must be nearby and may have
its parallax measured from the ground with a precision approaching
a milli arc-second.  According to Luyten (1957, 1979)
the number of high proper motion stars in the whole sky which are
brighter than $ \sim 15 $th magnitude is
$ N \approx 500 ~ (1''/pm)^3 $ (where $pm$ is the annual proper motion).
The scaling would be exact if all stars with a given proper motion
could be detected and measured, irrespective of their magnitude,
as for a given linear velocity the distance scales as $ pm^{-1} $,
and the volume and the number of stars scales as $ pm^{-3} $.  The
luminosity function is fairly flat at the faint end, so the number of high
proper motion stars brighter than 20th magnitude is likely to be only
a factor few larger than the numer of stars brighter than 15th magnitude.
There may be up to $ \sim 10 $ stars
which are brighter than 20th magnitude and which have
proper motion in excess of 0.5''/{\it yr} on a Schmidt plate
which is $ \sim 5 $ degrees on a side.

It will be very important to conduct a search for the faint high proper
motion stars over the whole sky.  When combined with the accurate
parallax measurements such a search would provide accurate information
about the local luminosity function at the faint end of the main
sequence.  The amount of work required by such a project is enormous,
and it is not clear when will it become feasible, though a tremendous
progress is being made in this area (cf. IAU Symposia 156 and 161:
Mueller and Kolaczek 1993, MacGillivray et al. 1994).
The purpose of this paper is to point out that once the very
laborious search for the faint, high proper motion stars is done,
and the objects are found, the follow-up determination of the masses
of a large subsample of such objects is relatively simple and
straightforward using gravitational microlensing.  This can be done
close to the Milky Way, where a high number density of distant stars
provides a convenient background for the microlensing searches.  In this
case we shall know ahead of time which are the lensing objects, and their
astrometry will make it possible to predict when the microlensing
events will occur.

Current CCD searches for gravitational microlensing allow the brightness
determination for up to 120,000 stars in a field of $ 15' \times 15' $
(Udalski {\it at al}. 1994), i.e. $ \sim 0.15 $ stars per square arc seconds.
This number depends on the seeing and can be made a factor few larger
under excellent seeing conditions.  For the purpose of this paper
we shall assume that the number of measurable stars may be as large
as $ 0.3 ~ (1'')^{-2} $.  Most of these stars are at a distance of
a few kiloparsecs (Paczy\'nski 1964, Paczy\'nski {\it et al.} 1994).

A high velocity star is likely to be at a distance of a few tens of
parsecs.  Let us consider it to be a lens with a mass $ M $ at a
distance $ D_d $, with one of the background stars being a source
at a distance $ D_s $.  The angular radius of the Einstein ring
can be calculated as (cf. Paczy\'nski 1991, eq. 4)
$$
\varphi _{_E} =
\left[ { 4 GM \over c^2} ~ { D_s - D_d \over D_d D_s } \right] ^{1/2}
= 0''.0090 \times
\left( { M \over 0.1 M_{\odot} } ~ { 10 ~ pc \over D_d } \right) ^{1/2} ~
\left( 1 - { D_d \over D_s } \right) ^{1/2} .  \eqno(1)
$$
Let the proper motion of the nearby star be $ \dot \varphi $.  The
area in the sky covered by it with its Einstein ring in time $ t $
can be calculated as
$$
S = 2 \varphi _{_E} \dot \varphi ~ t =
0.018 \times (1'')^2
\left( { M \over 0.1 M_{\odot} } ~ { 10 ~ pc \over D_d } \right) ^{1/2} ~
\left( 1 - { D_d \over D_s } \right) ^{1/2}
\left( { \dot \varphi \over 1'' ~ yr^{-1} } \right) ~
\left( { t \over 1 ~ yr } \right) .
\eqno(2)
$$
Combining the area given with the eq. (2) with the number density
of background stars estimated at 0.3 per $ (1'')^2 $ we find that
a typical time interval between microlensing events is somewhat
less than one per century for any given high proper motion star.
This means that we need at least a few hundred of them to detect
a few microlensing events per year.

The photometric measurements of microlensing induced time variations
give the value of the event time scale defined as
$$
t_0 \equiv { \varphi _{_E} \over \dot \varphi } , \eqno(3)
$$
and the mass of the lens follows directly from the eqs. (1) and (3) as
$$
M = 0.124 ~ M_{\odot}
\left( { \dot \varphi ~ t_0 \over 0''.01 } \right) ^2
\left( { D_d \over 10 ~ pc } \right)
\left( 1 - { D_d \over D_s } \right) ^{-1} .  \eqno(4)
$$
Note, that all the important quantities on the right hand side of eq. (4)
are directly measurable: $ \dot \varphi $, $ t_0 $, and $ D_d $.
The value of $ D_s $ is less important, as we expect $ D_d/D_s \sim
10^{-2} $, and even a crude estimate will be adequate.

The only technical problem with the microlensing is the difficulty of
achieving high photometric accuracy in dense stellar fields,
but this technology is rapidly maturing (cf. Udalski et al. 1994, Alcock
et al. 1995).  The proposed program is photometrically much easier
than the current massive microlensing searches, as the location and
time of each event can be predicted ahead of time, pretty much like
stellar occultations by asteroids.

The photometrically determined time scale of a microlensing event, $ t_0 $,
is sufficient to uniquely determine the mass of a single lensing star
or a brown dwarf.  Perhaps 1\% precision can be achieved, as
very accurate photometry (to obtain $ t_0 $), and astrometry (to obtain
$ D_d $) are possible, at least in principle.  If angular resolution
of $ \sim 0''.005 $ is achieved with the future imaging interferometers
then an independent astrometric determination of the size of the Einstein
ring, and therefore the lens mass, will be possible by means of precise
monitoring the relative location of the lens and the two micro-images,
which have the expected separation of the order of $ 0.''02 $, as given
with the eq. (1).

This project was supported with the NSF grants AST 92-16494 and  AST 93-13620.

\vskip 0.5cm
\centerline{\bf REFERENCES}
\vskip 0.5cm

\wc{Alcock, C. et al. 1995, {\it Phys. Rev. Letters}, {\bf 74}, 2867.  \hfill}

\wc{MacGillivray, H. T. et al. 1994, IAU Symp. 161: {\it Astronomy
from Wide Field Imaging}, Kluwer Academic Publiocations.  \hfill}

\wc{Mueller, I. I. and Kolaczek, B. 1993, IAU Symp. 156: {\it Developments
in Astrometry and Their Impact on Astrophysics and Geodynamics}, Kluwer
Academic Publications.  \hfill}

\wc{Paczy\'nski, B. 1964, {\it Acta Astr.}, {\bf 14}, 157. \hfill}

\wc{Paczy\'nski, B. 1991, {\it Astrophys. J. Letters}, {\bf 371}, L63. \hfill}

\wc{Paczy\'nski, B. 1994, {\it Acta Astr.}, {\bf 44}, 235. \hfill}

\wc{Paczy\'nski, B. Stanek, K.Z. Udalski, A., Szyma\'nski, M., Kaluzny, J.,
Kubiak, M., and Mateo, M. 1994, {\it Astron. J.}, {\bf 107}, 2060. \hfill}

\wc{Udalski, A. Szyma\'nski, M., Stanek, K.Z., Kaluzny, J., Kubiak, M.,
Mateo, M., Krzemi\'nski W., Paczy\'nski, B., and Venkat, R.  1994,
{\it Acta Astr.}, {\bf 44}, 165. \hfill}

\vskip 2.0cm

This paper has been submitted to Acta Astronomica on April 26, 1995

\vfill \end \bye